\documentclass[fleqn,10pt]{wlscirep}
\usepackage[utf8]{inputenc}
\usepackage[T1]{fontenc}
\usepackage[braket, qm]{qcircuit}
\usepackage{bm}
\usepackage{hyperref}

\title{Wavelet correlation noise analysis for qubit operation variable time series}

\author[1,2]{Amanda E. Seedhouse}
\author[1,2]{Nard Dumoulin Stuyck}
\author[1,2]{Santiago Serrano}
\author[1,2]{Will Gilbert}
\author[1]{Jonathan Yue Huang}
\author[1,2]{Fay E. Hudson}
\author[3]{Kohei M. Itoh}
\author[1,2]{Arne Laucht} 
\author[1,2]{Wee Han Lim}
\author[1,2]{Chih Hwan Yang}
\author[1,2]{Tuomo Tanttu}
\author[1,2]{Andrew S. Dzurak}
\author[1,2]{Andre Saraiva}

\affil[1]{School of Electrical Engineering and Telecommunications, The University of New South Wales, Sydney, NSW 2052, Australia}
\affil[2]{Diraq, Sydney, NSW, Australia}
\affil[3]{School of Fundamental Science and Technology, Keio University, Yokohama, Japan}

\begin{abstract}
In quantum computing, characterizing the full noise profile of qubits can aid in increasing coherence times and fidelities by developing error-mitigating techniques specific to the noise present. This characterization also supports efforts in advancing device fabrication to remove sources of noise. Qubit properties can be subject to non-trivial correlations in space and time, for example, spin qubits in MOS quantum dots are exposed to noise originating from the complex glassy behavior of two-level fluctuator ensembles. Engineering progress in spin qubit experiments generates large amounts of data, necessitating analysis techniques from fields experienced in managing large data sets. Fields such as astrophysics, finance, and climate science use wavelet-based methods to enhance their data analysis. Here, we propose and demonstrate wavelet-based analysis techniques to decompose signals into frequency and time components, enhancing our understanding of noise sources in qubit systems by identifying features at specific times. We apply the analysis to a state-of-the-art two-qubit experiment in a pair of SiMOS quantum dots with feedback applied to relevant operation variables. The observed correlations serve to identify common microscopic causes of noise, such as two-level fluctuators and hyperfine coupled nuclei, as well as to elucidate pathways for multi-qubit operation with more scalable feedback systems.
\end{abstract}

\begin{document}

\flushbottom
\maketitle

\thispagestyle{empty}

\section*{Introduction}

Electron solid-state devices are suitable for qubit systems due to their high tunability, but they are exposed to noise originating from the material stack. The noise acting on solid-state qubit systems can be complex, including non-Markovian noise~\cite{Burkard2009Non-MarkovianNoise,Chan2018AssessmentSpectroscopy,Morris2019Non-MarkovianIBMQX4}, and spatially and temporally correlated noise~\cite{Yoneda2022Noise-correlationSilicon}. In order to increase the fidelities of qubits so that error correction protocols can be implemented, noise should be well understood~\cite{Egan2021Fault-tolerantQubit,Abobeih2022Fault-tolerantProcessor,vanRiggelen2022PhaseQubits}. Besides quantum error correction, another approach called entanglement purification can distil high-quality entangled states from low-quality ones~\cite{Bennett1996Purification, kalb2017entanglement, OptExpress2020Purification, PhysRevA2022Purification, PhysRevLett2021Purification}. Both quantum error correction and entanglement distillation complement the strategies to understand qubit noise in this work.  Efforts in mitigating and measuring noise in qubit systems~\cite{Chan2018AssessmentSpectroscopy,Muhonen2014StoringDevice, Soare2014ExperimentalControl,Yoneda2018A99.9,Romach2015SpectroscopyDiamond,Bylander2011NoiseQubitb,Almog2016DynamicNoise,Morris2019Non-MarkovianIBMQX4}, including some focus on correlated noise~\cite{Szankowski2016SpectroscopyQubits,vonLupke2020Two-qubitQubits,Yoneda2022Noise-correlationSilicon,RojasArias2023Spatial} seek to understand the microscopic origin of noise by studying correlations between qubit operation variables~\cite{Yoneda2022Noise-correlationSilicon,RojasArias2023Spatial}. Building on this foundation, using qubits as a probe in solid-state devices has emerged as an effective strategy to study noise interactions.

In qubit devices, the dynamics and coherence of the system are perturbed by various sources of noise. The presence of correlations within the device operation variables is a result of the perturbations being a function of position and time. Temporal correlations manifest as a dependence on previous events, while spatial correlations on the location of the qubit within the device. The field of qubit error mitigation has made significant progress towards suppressing correlated noise contributions to qubit dynamics~\cite{seif2024suppressing}.

Spin qubits in silicon-based systems have achieved two-qubit gate fidelities above $99\%$~\cite{Mills2022Two-qubit99,Noiri2022AProcessors,Madzik2022PrecisionSilicon, Xue2022QuantumThreshold, Tanttu2023StabilitySilicon}, but there remains unexplained noise that presents itself within the $1\%$ infidelity. Improving error rates within this $1\%$ margin can significantly reduce the number of qubits required for error correction codes~\cite{Gidney2021HowQubits}. Additionally, a deeper understanding of noise can lead to customised codes for device-specific noise types~\cite{Ataides2021TheCode}, enhance control methods for quantum information processing~\cite{Yang2019SiliconEngineering,du2024error}, and improve insights into the materials used in device fabrication~\cite{saraiva2022materials}.

Therefore, having a tool that can handle and analyze large datasets for noise characteristics is crucial. Recently, the amount of data that can be collected from spin qubit systems has vastly increased due to the development of real-time logic with controllers supported by field-programmable gate arrays (FPGAs). With vast amounts of data being produced, it is appropriate to move to more suitable statistical techniques. To do this, we investigate techniques from scientific communities that already manage large data sets. Wavelet analysis has enhanced weather predictions~\cite{Jiang2020RefiningModeling, Jiang2021AAnomalies}, trading forecasts~\cite{AlWadi2011SelectingModel}, and traffic accuracy~\cite{Li2023TrafficTransformation} by focusing on specific time scales. In quantum related fields, wavelet analysis has been used to improve the reading out of charge jumps in a solid-state device~\cite{Prance2015IdentifyingDetection}, and spin dynamics studies based on ensembles~\cite{Phillies1995WaveletDynamics,Guldeste2022WaveletEnsemble} and nitrogen vacancy centres~\cite{guldeste2023waveletbased}. Using wavelets, signals can be decomposed in a basis of a wave localised in time~\cite{Percival2000WaveletAnalysis}, resulting in frequency information about the signal, as well as non-periodic time information. This is in contrast to Fourier-based analysis which is specifically useful for identifying particular periodicities in data.

The present study employs wavelet-based analysis on an example data set from a qubit system using a silicon metal-oxide-semiconductor (SiMOS) device comprising two qubits. We analyze feedback data performed with the aid of an FPGA tracking eight different single- and two-qubit operation variables in a long ($>3\times10^4$ s) experiment~\cite{DumoulinStuyck2023FeedbackExperiment}. The data contains information on noise that couples to the qubits directly. Wavelet-based analysis is used to extract information on the noise which affects the feedback operation variables through the wavelet transformation, as well as to extract correlations through the coherence function, the Pearson correlation coefficient, and the wavelet variance transformation~\cite{Jiang2020RefiningModeling}. We show that the analysis is sensitive to the chosen basis. This paper uses wavelet-based analysis on vast quantities of data to comprehend noise in solid-state qubit systems, showing their potential to explore noise analysis in different qubit systems. 

\begin{figure}
    \centering
    \includegraphics[width=1\linewidth]{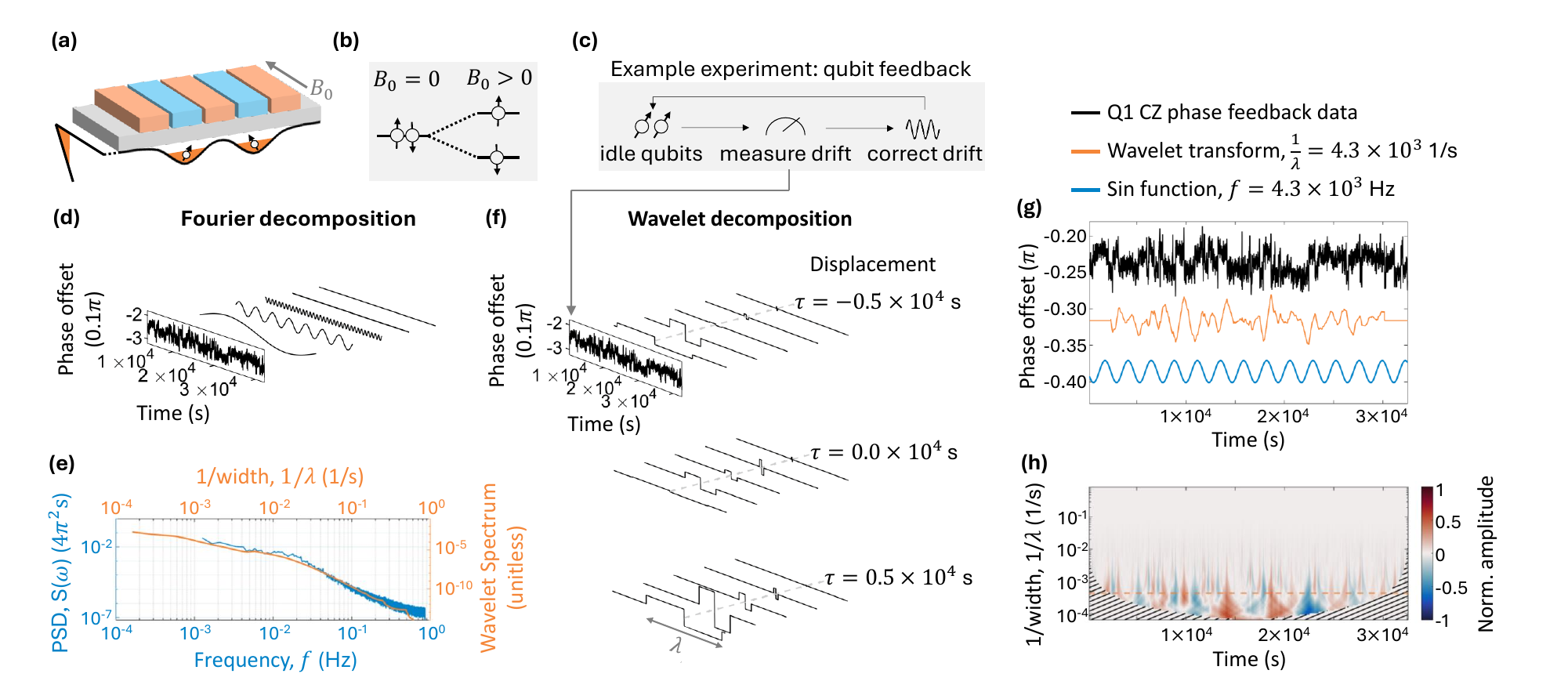}
    \caption{(a) Schematic of the device used in the experiment, including the potential profile (black) created in the device. (b) The effect of Zeeman splitting the spin up and spin down states with an external $B_0$ field. (c) A schematic of the experimental protocol. (d) Fourier analysis of Q1 phase feedback data. On the left is a plot of the data; on the right is an inset of the power spectral density of the data, and in the middle is the data decomposed into select frequencies. Above this is a simple representation of the experiment. (e) The power spectral density in blue and the wavelet spectrum in red. (f) Wavelet analysis using the Haar wavelet of Q1 phase feedback data. On the left is a plot of the feedback data; on the right is an inset of the wavelet transformation of the data normalised to the maximum value for displacements $\tau = -0.5 \times 10^{4} \, \text{s}, 0 \, \text{s},$ and $0.5 \times 10^{4} \, \text{s}$, and in the middle is the data decomposed into select wavelet widths $\lambda$. (g) The full Q1 phase feedback data set (black) decomposed into the Haar wavelet transformation at $\lambda = 2.35 \times 10^{3} \, \text{s}$ (red) and the Fourier component of the data at $f = 4.26 \times 10^{-4} \, \text{Hz}$ (blue). The wavelet transformation and Fourier component are scaled and displaced for ease of reading with respect to the data set. (h) The full wavelet transformation of the data normalised to the maximum value. The lined section indicates the cone of influence.
}
    \label{fig:wavelet_intro}
\end{figure}

\section*{Example data set}
While the methods discussed in this work are universally applicable to time series, here we specifically focus on analysing an experiment performed in Ref.~\cite{DumoulinStuyck2023FeedbackExperiment}, which carries out feedback protocols on qubit operation variables in which we continuously monitor the qubits. This section introduces the experiment.

The qubits reside in a silicon substrate below a silicon dioxide interface, bound by the electric field created by aluminium metal gates deposited on top of the oxide, giving rise to a triangular-shaped potential with its trough at the junction between the silicon and its oxide layer~\cite{Ando1982ElectronicSystems}, a schematic of the potential landscape of the device is given in Fig.~\ref{fig:wavelet_intro}(a). The potential arises from the differences in the electronic properties of silicon and silicon dioxide. The shape and height of the potential can be tuned by the voltage applied to the gate. This modulation of the potential allows for control of the electron density in the silicon channel; multiple gate electrodes enable further tuning of the potential landscape resulting in quantum dots that contain a controlled number of electrons. The device is tuned to a regime where two quantum dots are formed. We define qubit 1 (Q1) as the left dot and qubit 2 (Q2) as the right dot written as (1,3), denoting the number of electrons in (Q1, Q2). In our case, three electrons reside in one of the dots where the two lowest energy electrons form a closed shell and do not interfere with the spin dynamics of the other electron that is used as the qubit~\cite{Tanttu2023StabilitySilicon,Seedhouse2021PauliControl, Yang2020OperationKelvin, Leon2021Bell-stateMolecule}. The other dot contains a single electron. More details on the device architecture can be found in Refs.~\cite{Tanttu2023StabilitySilicon, DumoulinStuyck2023FeedbackExperiment}. 

The single electron spin states are Zeeman split by an in-plane magnetic field $B_0$, illustrated in Fig.~\ref{fig:wavelet_intro}(b), defining the Larmor frequency $g\mu_BB_0$ of each two-level system that can be used as a qubit ($g$ is the electron g-factor and $\mu_B$ is the Bohr magneton). The Larmor frequency can be determined using a Ramsey experiment~\cite{Vandersypen2004NMRComputation}. In this experiment, the electron spin is aligned perpendicular to the $B_0$ field, causing the electron to precess about $B_0$ at the Larmor frequency. This setup allows us to measure the frequency of spin precession by observing the measured projection along the $B_0$ axis change as a function of time. To perform single-qubit gates, an alternating magnetic field $B_1$, oriented perpendicular to $B_0$, is generated via an on-chip antenna by applying microwave pulses. Transforming into the rotating frame set by the frequency of the microwave source~\cite{Vandersypen2004NMRComputation}, the electron spin precesses about the, now stationary, $B_1$ field causing Rabi oscillations~\cite{Veldhorst2015ASilicon} when the frequency of the microwave is resonant with the qubit Larmor frequency. The Rabi frequency scales proportional to the square root of the applied microwave power. Two-qubit interactions are realized through the exchange interaction present when there is an overlap between the two electron wavefunctions~\cite{Tanttu2023StabilitySilicon, Meunier2011EfficientDots}. The exchange interaction is controlled through gate electrodes that are placed between the dot forming electrodes. The voltage applied to the interstitial gate tunes the magnitude of the exchange interaction. To readout the state of the qubits, a Pauli spin blockade readout method called parity readout is implemented, which uses a single electron transistor to probe the charge movement of the electrons. Parity readout results in the single electron transistor detecting a charge movement from (1,3) to (0,4) if the spin states of the two electrons are anti-parallel (i.e. in a singlet state or $T_0$ triplet state), otherwise there is no charge movement~\cite{Yang2020OperationKelvin,Seedhouse2021PauliControl}.

The data analyzed in this work tracked eight operation variables in a feedback experiment consisting of a two-qubit system left to shift over time, away from its calibrated values. The total experiment is $>3\times10^4$ s long, with between 55,500-55,600 feedback steps (depending on the variable) consisting of a measurement followed by a correction so that each variable does not drift too far. This method of feedback is used to allow for long experiments to be performed on qubit systems without the whole system having to be re-calibrated when a large drift occurs~\cite{Tanttu2023StabilitySilicon}. Here, a full feedback protocol can take up to 0.57 s (consisting of approximately 20 repeats of the quantum circuit and measurement) setting the smallest time-scale of noise we can analyze. The variables tracked are: the microwave pulse power of both qubits, corresponding to the Rabi frequencies; the Larmor frequencies; the individual qubit phases acquired by pulsing the exchange control gate due to Stark shift~\cite{Tanttu2023StabilitySilicon,watson2018programmable}, which we call qubit CZ phases to correspond to the implementation of a controlled-Z gate; the voltage applied to the exchange control gate to obtain a target exchange interaction, which we call exchange level for short; and the (1,3)-(0,4) charge transition voltage level, corresponding to the readout point. Each variable is measured, recorded, then corrected compared to an initial calibration test where the offset in the feedback is set to zero, with Fig.~\ref{fig:wavelet_intro}(c) illustrating this protocol. Some variables' initial calibration were not on target, as evidenced by a significant drift in the first 200 seconds of the data (cut off in Figs.~\ref{fig:wavelet_intro} and \ref{fig:variancetransform}) demonstrating the capabilities of the feedback. More details on how these were measured and the full data sets can be found in Ref.~\cite{DumoulinStuyck2023FeedbackExperiment}. 

This device, along with similar platforms, provides a valuable environment for studying noise, thanks to a well-established field of research that has already identified several major sources of noise in this hardware. A predominant noise source in SiMOS qubits is low-frequency $1/f$ noise, attributed to a collective of two-level fluctuators~\cite{Machlup1954NoiseSignal, Hooge2003OnNoise}. There may also be prominent fluctuators within the ensemble impacting the qubit environment~\cite{Elsayed2022LowManufacturing}. Fluctuators can magnetically or electrically couple to an electron spin qubit via spin-orbit interaction, with nuclear spins being candidates for magnetic coupling. The device~\cite{DumoulinStuyck2023FeedbackExperiment}, constructed from isotopically enriched silicon with 800ppm $^{29}$Si nuclei, implies a significant likelihood of hyperfine coupling between $^{29}$Si nuclei and electron qubits~\cite{Hensen2020AQubit}. A flip in a hyperfine-coupled $^{29}$Si nucleus can alter the qubit Larmor frequency. Electrically, fluctuators are likely trapped charges, tunnelling into defects within the device~\cite{Hooge2003OnNoise}. These microscopic noise sources are distinguishable since (i) hyperfine interactions are contact interactions, affecting only electrons in direct contact, and (ii) electrically coupled fluctuators also influence operation variables sensitive to the electrical environment, unlike the magnetic, e.g., the exchange and readout point. By understanding these noise sources, we can refine our analysis methods, align them with known sources, and identify features in data sets that may or may not correspond to these categories.

\section*{Analysis methods}
In this section, we introduce wavelet and Fourier signal analysis techniques. Fourier analysis, which decomposes signals into sinusoids, is well-suited for stationary, periodic signals but lacks time localization, making it less effective for transient behaviors. In the context of this experiment, Fourier analysis could effectively capture the decomposed frequencies of periodic signals influencing the qubit parameters. Wavelet analysis, however, provides greater flexibility, offering high resolution for signals with sudden changes or varying frequencies, such as those found in a discrete two-level system jumps in this device. Additionally, wavelet analysis allows to capture transient behaviour of the periodic signal.

\subsection*{Wavelet transformation}

Wavelets are mathematical functions that are a useful tool for the analysis of time series~\cite{Percival2000WaveletAnalysis,Jiang2020RefiningModeling,Prance2015IdentifyingDetection} due to the ability to decompose data sets into time and frequency components. This work focuses on the use of the continuous wavelet transformation that is discretized to enable the analysis of discrete data sets. The process of discritization is shown in the Methods section. We define the mother wavelet to be a unitless vector $\vec{\Psi}=(\Psi_0, \Psi_1, ..., \Psi_{N-1})^T$ where $(T)$ is the transpose. The two basic properties of a wavelet are
\begin{equation}
    \sum_{n=0}^{N-1}\Psi_n=0
\hspace{0.5cm}\text{and}\hspace{0.5cm}
    \sum_{n=0}^{N-1}|\Psi_n|^2=1.
    \label{eq:waveletconditions}
\end{equation}
One example of a type of wavelet is the Haar wavelet~\cite{Haar1911ZurFunktionensysteme}, depicted in the time domain in Fig.~\ref{fig:wavelet_intro}(f).

The mother wavelet can be displaced by $m = \{0, ..., N - 1\}$ and scaled with $ k = \{2, 4, 6, ..., K\}$ such that a wavelet reads $\psi_n(m,k) = \psi[(n-m)/k]$. The discrete Haar wavelet function is given as
\begin{equation}
\psi^{\textrm{H}}\bigg(\frac{n-m}{k}\bigg) = \psi^{\textrm{H}}_n(m,k) = \left\{
    \begin{array}{ll}
        \frac{1}{2\sqrt{k}}, \hspace{0.8cm} m-k/2 \leq n \leq m; \\ \\
        -\frac{1}{2\sqrt{k}}, \hspace{0.5cm} m< n \leq m+k/2; \\ \\
        0, \hspace{1.25cm} \text{otherwise}.
    \end{array}
\right.    
\end{equation}
From this definition, all non-zero components of the Haar wavelet cover a time of $k$. 

Another type of wavelet that is explored in this work is the Morlet wavelet, defined as a plane wave confined by a Gaussian window and discretized so that
\begin{equation}
    \psi^{\textrm{M}}_n(m,k) = \frac{1}{\sqrt{|k|}}\left( e^{-i \epsilon \frac{n - m}{k}} - e^{-\frac{\epsilon^2}{2}}\right) e^{-\frac{1}{2}\left(\frac{n - m}{k}\right)^2},
\end{equation}
with $\epsilon$ being the parameter that scales the Gaussian window. This allows one to tune between capturing more information in frequency or in time. The term $e^{-\frac{\epsilon^2}{2}}$ ensures the conditions in Eqs.~\ref{eq:waveletconditions} are satisfied.

The Haar wavelet is ideal for detecting sharp, discrete changes in signals, making it suitable for noise or abrupt shifts, such as those in electrical fluctuations or two-level systems known to play an important role in spin qubit systems~\cite{Elsayed2022LowManufacturing}. Its simplicity captures binary-like structures. In contrast, the Morlet wavelet excels at analyzing smooth, oscillatory signals with varying frequencies, offering higher resolution in the frequency domain. It balances time and frequency localization, making it effective for gradual transitions. Together, these wavelets allow a broader understanding of qubit noise, capturing both discrete jumps and smooth oscillations.

The wavelet transform can analyze a time series $\vec{x} = (x_0, x_1, ..., x_{N-1})^T$~\cite{Torrence1998AAnalysis}. The discretized continuous wavelet transform applied to a discrete time series is expressed as
\begin{equation}
W(m,k) = \sum_{n=0}^{N-1} x_n \psi_n^*(m,k),
\label{eq:wavelet_transform}
\end{equation}
where $(*)$ is the complex conjugate. More discussion on the discretization of the wavelet transformation and the properties of the wavelet are discussed in the methods section.

A wavelet transformation on $\vec{x}$ that runs over values of $m$ and values of $k$, results in the matrix 
\begin{equation}
\bm{W}_x = 
\begin{pmatrix}
W(0,2) && W(0,4) &&\hdots &&W(0,K)\\
W(1,2) && W(1,4) &&\hdots &&W(1,K)\\
\vdots &&\ddots &&\ddots &&\vdots \\
W(N-1,2) && W(N-1,4) &&\hdots &&W(N-1,K)
\label{eq:wavelet_matrix}
\end{pmatrix}
\end{equation}
with dimensions $N \times (N-1)\ \mathbf{div}\ 2$ (with $\mathbf{div}$ denoting integer division) and in the same units as the variable $\vec{x}$. This example of the wavelet transformation runs over all possible $m$ and $k$ values rendering the transformation over-complete. A theory of discrete wavelets can remove the redundancies~\cite{Daubechies1988OrthonormalWavelets}, however in our analysis of noise we retain all information.

To maintain a link to the time domain, we define time as $t=n\Delta t$ with an interval of $\Delta t$, the time displacement $\tau=m\Delta t$ and translation $\lambda=k \Delta t$ with time units. The frequency content of the wavelet function is inversely related to the width $\lambda$. A larger $\lambda$ corresponds to a more stretched wavelet that can capture lower-frequency information, while a smaller $\lambda$ corresponds to a more compressed wavelet that can capture higher-frequency information. While $1/\lambda$ is not directly the frequency, it is analogous to frequency in that it is inversely related to the scale at which the function analyzes the signal. The wavelet transformation is analogous to the windowed Fourier transform~\cite{Daubechies1992TenWavelets}, however the wavelet transformation allows for finer resolution when analysing high frequency signals that appear for small time steps. This is due to the fact that $\lambda$ can be changed, while the windowed Fourier transform has a fixed window size.

\subsection*{Correlation techniques}
The standard approach in correlation analysis is given by the cross-power spectral density, defined as
\begin{equation}
S_{xy}(f) = \frac{1}{N} X(f) \cdot Y^*(f),
\end{equation}
where $f$ is frequency, $X(f) = \sum_{n=0}^{N-1} x_n \cdot e^{-i2\pi fn\Delta t}$ and $Y(f) = \sum_{n=0}^{N-1} y_n \cdot e^{-i2\pi fn \Delta t}$. Here, $X(f)$ and $Y(f)$ are the Fourier transforms of the signals $x_n$ and $y_n$, representing how much power of each signal is present at frequency $f$. The cross-power spectral density $S_{xy}(f)$ measures how much two signals $x_n$ and $y_n$ co-vary as a function of frequency. To remove features that occur due to the finite and discrete nature of sampled data, power spectral densities (PSDs) are typically smoothed; in this work, we use the Welch method~\cite{welch1967use} for smoothing.

The cross-PSD can be normalised with the auto-power spectral densities $S_{xx}(f) = \frac{1}{N} X(f) \cdot X^*(f)$ and $S_{yy}(f) = \frac{1}{N} Y(f) \cdot Y^*(f)$, which represent the power distributions of individual signals across frequencies. This normalisation results in the coherence function
\begin{equation}
C_{xy}(f) = \frac{|S_{xy}(f)|^2}{S_{xx}(f)S_{yy}(f)},
\label{eq:norm_cross_psd}
\end{equation}
which provides a frequency spectrum of linear correlations, where a value of 0 indicates no correlation and 1 signifies high correlation.

A similar analysis can be applied using the wavelet transformation; the cross-wavelet transformation is defined by $\bm{W}_{xy} = \bm{W}_x\bm{W}^{*}_{y}$ and can be normalised by the auto-wavelet transformation defined in Eq.~\ref{eq:wavelet_transform} to yield the wavelet coherence 
\begin{equation}
    \bm{C}_{xy} = \frac{s(\bm{W}_{xy})}{s(\bm{W}_x) s(\bm{W}_y)},
    \label{eq:wavelet_coherence}
\end{equation}
where the function $s$ represents a smoothing operation over time and scale. In the absence of a smoothing function, calculated correlations between the two signals are not reliable. The ideal choice of smoothing function varies with the wavelet basis used; for the Morlet wavelet, time smoothing is performed with a Gaussian $e^{-t^2/(2\tau^2)}$, and scale smoothing with a boxcar filter of width 0.6, as detailed in Ref.~\cite{torrence1999interdecadal}. The sensitivity of wavelet coherence to the chosen smoothing function is comparable to that of Fourier coherence. Given the lack of a natural smoothing function for the Haar wavelet in the literature, our analysis uses only the Morlet wavelet coherence function. This function provides both frequency and temporal information on correlations, where a value of 1 indicates perfect correlation and 0 indicates no correlation.

\subsection*{Wavelet variance}
The wavelet variance is a useful tool in recognising significant time scales within datasets because it uses the variances in each dataset to enhance features in one another by identifying prominent features with a certain width $\lambda$. We refer to the variables as the \textit{predictor}, $\vec{x}$, and the \textit{response}, $\vec{y}$. The variance transformation is introduced in Ref.~\cite{Jiang2020RefiningModeling} as a means to identify a meaningful predictor variable and formulate a predictive model between $\vec{x}$ and $\vec{y}$. The former identification step is our focus when applying to quantum systems, although the predictive model can be explored in future research. The terms predictor and response are used because the predictor variable is modified to improve the accuracy of forecasting the behavior of the response variable.

The variance transform compares the response $\vec{y}$ to the wavelet transformation $\bm{W}_x$ of the predictor variable $\vec{x}$. This is done by calculating the covariance between $\vec{y}$ and $\bm{W}_x$. The result is a vector of covariances $\vec{C}=(C_1, C_2,...,C_K)^T$ that compares $\vec{y}$ to the wavelet transformation at a given $\lambda$, i.e.,
\begin{equation}
    \vec{C}^T = \frac{(\vec{y}-\bar{\vec{y}})^T \bm{W}_{x}}{N-1},
\end{equation}
where $\bar{\vec{y}}=(\bar{y}, \bar{y},...,\bar{y})^T$ is a vector of length $N$, with each entry being the mean $\bar{y} = \sum_{n=0}^{N-1}y_n/N$. The variance-transformed data is then
\begin{equation}
    \vec{x}' = \bm{W}_x \sigma_x \vec{C},
    \label{eq:variance}
\end{equation}
where $\sigma_x$ is the standard deviation of $\vec{x}$. The variance transformation redistributes the variances for each $\lambda$ component in $\vec{x}$ according to $\vec{C}$. The result is a new time series $\vec{x}'$ that exhibits similar spectral properties as $\vec{y}$, the key to this technique.

For the variance transformation method to be viable, all datasets being compared must have the same time step. To achieve this, the data is formatted to have the same time step as the data with the largest step (in our case, this is the Q1 Larmor frequency feedback), which involves removing intermediate data points. Furthermore, as $1/\lambda$ decreases, the cone of influence increases, and therefore there is a limit to the minimum $1/\lambda$ in $\bm{W}_x$ that can be used in the variance transformation. A cut-off $1/\lambda$ is chosen as the minimum value such that $80\%$ of the data remains trustworthy. 

\section*{Results}

\subsection*{Wavelet transformation on qubit data}
The wavelet transform in Eq.~\ref{eq:wavelet_matrix}, through convolution with a time series, assesses similarities across inverse withs $1/\lambda$. Fig.~\ref{fig:wavelet_intro} shows Fourier and wavelet analyzes of the process for the phase acquired by pulsing the exchange control gate for Q1. Fourier decomposition into frequency $f$ components is depicted in Fig.~\ref{fig:wavelet_intro}(d), while wavelet analysis, focusing on inverse width values ($1/\lambda$) and time shifts ($\tau$), is shown in Fig.~\ref{fig:wavelet_intro}(f). Fourier analysis excels at highlighting how signal power varies with frequency, $f$, useful for detecting periodic signals, as demonstrated by the power spectral density inset in Fig.~\ref{fig:wavelet_intro}(e).

Wavelet transformation offers insights into the signal's spectral information, akin to Fourier's power spectral density, by calculating the variance $\sigma^2$ of $\bm{W}_x$ for each $1/\lambda$, leading to the wavelet spectrum $\sigma^4$. This comparison between Fourier and wavelet spectra, shown in Fig.~\ref{fig:wavelet_intro}(e), reveals qualitative agreement, although Fourier captures finer frequency details because it uses sinusoids that extend over the entire time domain. 

Fig.~\ref{fig:wavelet_intro}(f) presents $\bm{W}_x$ normalised to maximum value, for different $\tau$ values, visually representing data decomposition into the wavelet basis. This is further explored in Fig.~\ref{fig:wavelet_intro}(h), displaying the full $1/\lambda$ and $\tau$ time series representation. The wavelet transformation's ability to detect discrete transitions in data is visualised in Fig.~\ref{fig:wavelet_intro}(g), aligning peaks and troughs with data transitions. The choice of different wavelet basis, like the Haar for identifying discrete jumps as seen in Fig~\ref{fig:wavelet_intro}(g), emphasises different system characteristics, aiding in event identification and enhancing signal correlation analysis.

A limitation associated with the dataset size is that the range of $1/\lambda$ values, into which a given wavelet can decompose the signal, is limited by the requirement for the wavelet to decay to zero within the signal's duration. This necessitates conducting experiments over a duration long enough relative to the scale of features under investigation. Moreover, boundary effects, known as the cone of influence~\cite{Percival2000WaveletAnalysis}, must be taken into account. This cone, depicted in Fig.~\ref{fig:wavelet_intro}(h) as a lined area expanding as $1/\lambda$ decreases, indicates which portions of the wavelet transformation are reliable, ensuring wavelets at the transformation's edges have sufficiently decayed. While boundary effects can limit the precision of wavelet analysis at the signal’s edges, the method's flexibility in choosing time and frequency resolutions allows to compensate for this, providing insights into transient effects in data.

Utilizing the wavelet transformation, we investigate the microscopic origin of noise affecting the operation variables. Spectral analysis, facilitated by wavelets~\cite{Wornell1993Wavelet-BasedProcesses}, offers an intuitive examination of $1/f$ noise, as demonstrated in Fig.~\ref{fig:wavelet_intro}(e) for $1/\lambda = 0.3\times 10^{-4}$ s$^{-1}$ to $10^{-2}$ s$^{-1}$. The wavelet spectrum reveals a notable bump at $1/\lambda = 10^{-2}$ s$^{-1}$ with a $\lambda^2$ decline, suggestive of a prominent fluctuator within the ensemble impacting the qubit environment~\cite{Elsayed2022LowManufacturing}.

\begin{figure}
    \centering
    \includegraphics[width=1\columnwidth]{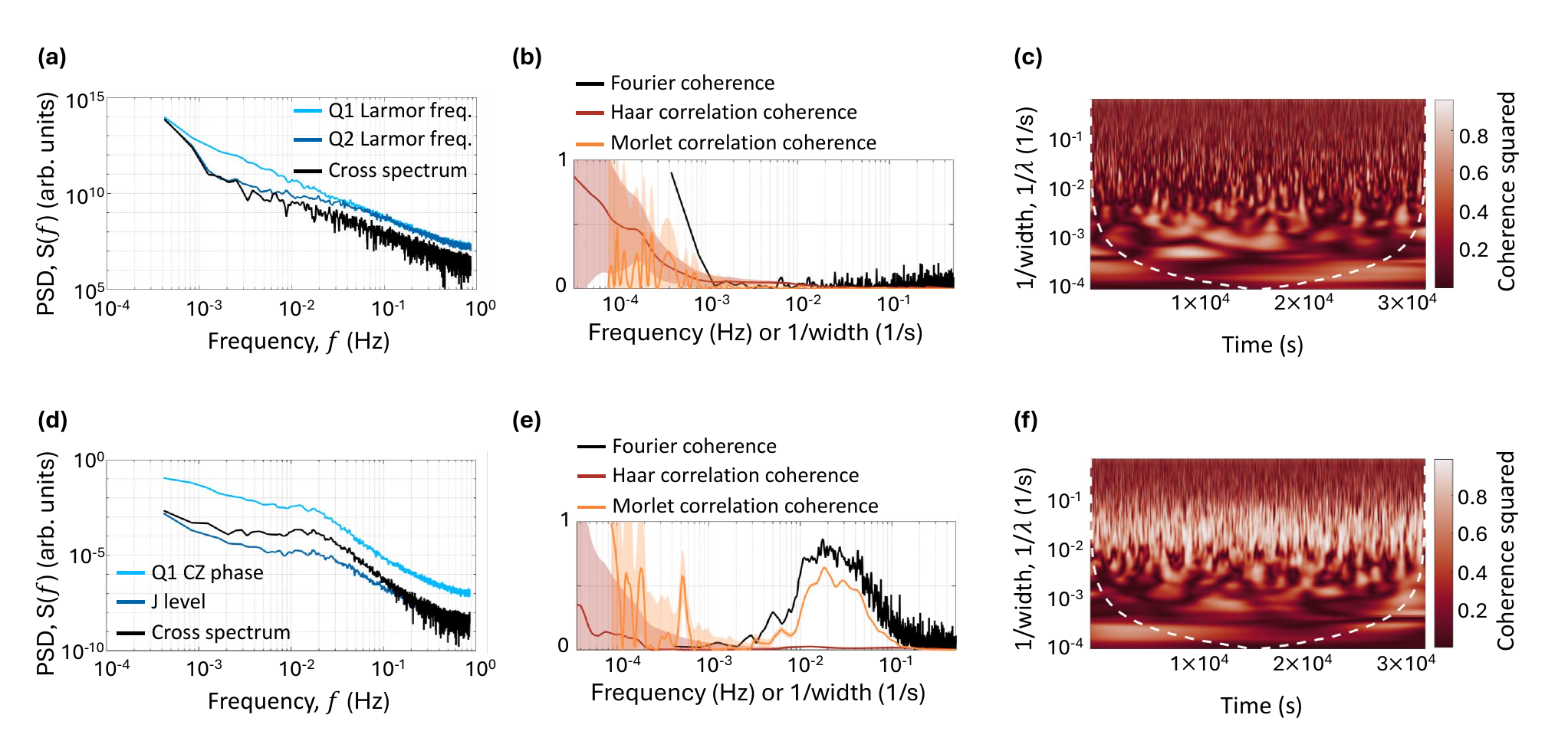}
    \caption{Fourier and wavelet correlation analysis for the Q1 Larmor frequency and the Q2 Larmor frequency data (a)-(c): (a) The auto (blue) and cross (black) power spectral densities calculated for Q1 Larmor frequency and the Q2 Larmor frequency. (b) Normalised cross power spectral density defined in Eq.~\ref{eq:norm_cross_psd} and the Pearson correlation coefficient for Q1 Larmor frequency and the Q2 Larmor frequency using the Haar (dark orange) and Morlet (light orange) basis. The shaded background shows the confidence in the value, calculated as the percentage of data within the cone of influence. (c) The normalised cross wavelet transformation in the Morlet basis. Q1 CZ and  J level feedback data (d)-(f): (d) Auto (blue) and cross (black) power spectral densities calculated for Q1 CZ and  J level analysis. (e) Normalised cross power spectral density defined in Eq.~\ref{eq:norm_cross_psd} and the Pearson correlation coefficient for Q1 CZ and  J level analysis using the Haar (dark orange) and Morlet (light orange) basis. (f) Normalised cross wavelet transformation in the Morlet basis.
}
    \label{fig:wavelet_coherence}
\end{figure}

\begin{figure}
    \centering
    \includegraphics[width=1\columnwidth]{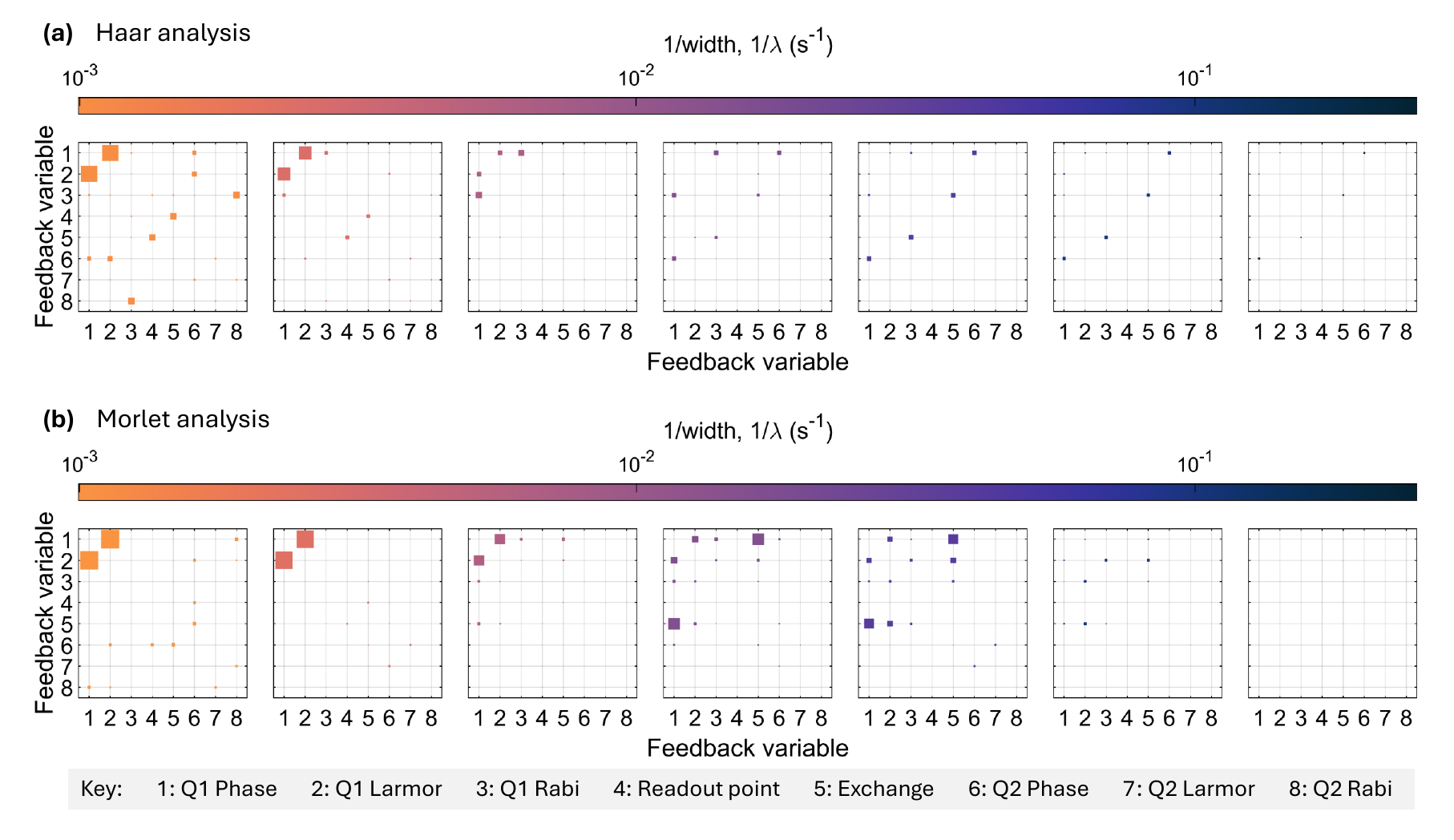}
    \caption{The $r^2$-value for a range of $1/\lambda$ values in (a) the Haar basis and (b) the Morlet basis. The key at the bottom indicates the feedback variable. In these plots, the size of the square corresponds to the square of the Pearson correlation coefficient, $r^2$-value, and the colour corresponds to $1/\lambda$. 
}
    \label{fig:correlation}
\end{figure}

\subsection*{Correlation Analysis on qubit data}
Our investigation reveals the wavelet coherence function's ability to uncover correlations over distinct time intervals, as highlighted in Fig.~\ref{fig:wavelet_coherence} using Eq.~\ref{eq:wavelet_coherence}. An initial analysis of the Q1 and 2 Larmor frequencies uses Fourier analysis, plotting both the cross and auto power spectral densities in  Fig.~\ref{fig:wavelet_coherence}(a), with the Fourier coherence in  Fig.~\ref{fig:wavelet_coherence}(b), found using Eq.~\ref{eq:norm_cross_psd}. Wavelet analysis is performed as a function of only the wavelet width. The Pearson correlation coefficient, denoted as the $r^2$-value, is calculated for each data pair across the widths of their wavelet transformations. These calculations are plotted in Fig.~\ref{fig:wavelet_coherence}(b) for both the Haar and Morlet wavelet bases. Together, these suggest minimal correlation across the spectrum. At lower frequencies, sampling errors start to dominate; hence, our analysis discussions are limited to frequencies above $10^{-3}$ Hz. This restriction means we do not address the large time-scale correlations~\cite{DumoulinStuyck2023FeedbackExperiment}, which are attributable to the drift of the superconducting magnet from the experimental setup. Nonetheless, the Morlet wavelet coherence, detailed in Fig.~\ref{fig:wavelet_coherence}(c), reveals time-specific correlations that Fourier and wavelet frequency-only analysis fails to discern. For example, at $t=1\times10^4$s and 1/$\lambda=10^{-3}$ 1/s there is a distinct peak in wavelet coherence squared is observed on the 2D coherence map. This peak indicates a transient event that occurs in both Larmor frequency data sets, but only for a brief duration. Such features are particularly valuable as they can reveal correlations between the datasets that are not persistent over time. Unlike traditional Fourier analysis, which primarily provides a global view of frequency content averaged over the entire signal duration, wavelet analysis maintains the temporal context. This allows for the identification of events that occur at specific times and frequencies, which are typically averaged away when considering only frequency information. 

When examining the Q1 CZ phase and J level data sets, a pronounced correlation at approximately $0.02$ Hz is observed throughout the time series in Figs.~\ref{fig:wavelet_coherence}(d)-(f). This persistent periodic correlation, captured by both the Morlet wavelet and Fourier analyzes but not by the Haar wavelet, suggests the underlying correlation's source might escape detection by methods not suited for slowly varying features therefore revealing the functionality of the correlation. The CZ phase of Q1 can be impacted by both hyperfine interactions and charge noise, whereas the J-level is affected solely by charge noise. This suggests that the primary origin of the noise is likely electrical. The findings presented in Fig.~\ref{fig:wavelet_coherence}(f) are intriguing; if the noise source were an electrically coupled single two-level fluctuator, one might anticipate the Haar wavelet to detect the noise correlation. Consequently, it appears unlikely that this correlation feature originates from a two-level fluctuator. Further work would need to be done to pinpoint the noise origin.

We perform further analysis encompassing eight variables from the feedback experiment~\cite{DumoulinStuyck2023FeedbackExperiment} by calculating the $r^2$-value for each wavelet transformation width for all pairs of data. Fig.~\ref{fig:correlation}(a) shows this analysis using the Haar wavelet basis, and Fig.~\ref{fig:correlation}(b) the Morlet basis. The variables are arranged such that the intra-qubit temporal correlations are located in the top left and bottom right corners, while the inter-qubit temporal and spatial correlations are found in the top right and bottom left corners. The diagonal terms are not calculated since they are comparing each data set to itself.

The Haar wavelet is suited for identifying discrete transitions, so it is probable that correlations in Fig.~\ref{fig:correlation}(a) originate from electrical or hyperfine-coupled two-level fluctuators, as inferred from the physics of the SiMOS device~\cite{saraiva2022materials}. For instance, examining the Q1 phase and Q1 Larmor frequency reveals a notable high magnitude correlation at $1/\lambda < 10^{-2}$ s$^{-1}$. Fluctuation signatures present in the raw datasets--see Fig.~\ref{fig:variancetransform}(a) for an example--occur at various time scales, corresponding to the significant correlations at $1/\lambda < 10^{-2}$ s$^{-1}$. A correlation between the Q1 Larmor and Q1 phase is expected due to a shift in the Larmor frequency inducing a phase alteration relative to the rotating frame, inherently linked to the rate of the electron spin's precession about the $B_0$ axis. Notably, no other variables show a high correlation with either the Q1 phase or Q1 Larmor, indicating an absence of similar characteristics at this timescale among other variables. This observation, along with discrete jumps in the data in Fig.~\ref{fig:variancetransform}(a) and a spectral bump in Fig.~\ref{fig:wavelet_intro}(b), supports the notion of a nucleus hyperfine-coupled to Q1. The hypothesis excludes an electrically coupled two-level fluctuator as the source considering such a fluctuator has the ability to interact with both Q1 and Q2 by shifting the electrical environment within the device. Further experiments are required to conclusively determine the nature of these interactions.

The Morlet wavelet analysis, as presented in Fig.~\ref{fig:correlation}(b), exhibits different characteristics due to the Morlet wavelet's capability in detecting slow-transitioning features within data. As previously discussed, the Q1 CZ phase and J level datasets demonstrate significant correlation, which is evident in the figure for $1/\lambda$ values ranging between $1\times10^{-2}$ s$^{-1}$ and $1\times10^{-1}$ s$^{-1}$. The Morlet analysis identifies correlations between the Q1 phase and Q1 Larmor frequency at the same $1/\lambda$ values as observed with the Haar basis (less than $1\times10^{-2}$ s$^{-1}$). However, the correlation strength differs, with the Morlet basis indicating stronger correlations. This suggests that specific features within the datasets, correlated at these scales, are effectively captured by both the Haar and Morlet bases, though with differing levels of intensity.

\subsection*{Variance Transformation on qubit data}
\begin{figure}
    \centering
    \includegraphics[width=\columnwidth]{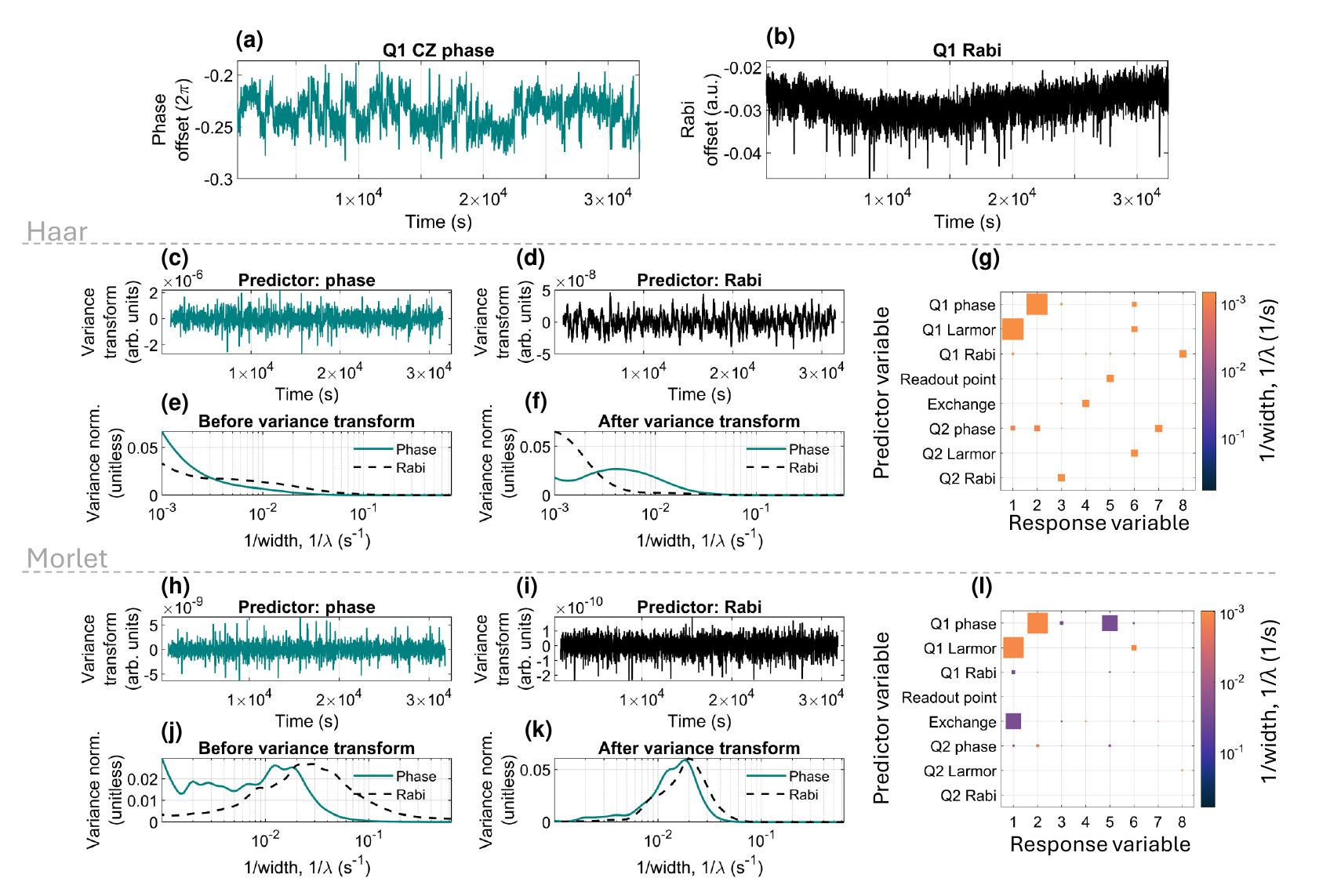}
    \caption{(a) Q1 phase feedback data, (b) Q1 Rabi feedback data. Both (a) and (b) start at an offset of zero, but this data is not shown; the full data sets can be found in Ref.~\cite{DumoulinStuyck2023FeedbackExperiment}. The variance-transformed data sets found using Eq.~\ref{eq:variance} with (c) the Q1 phase data as the predictor variable and (d) the Q1 Rabi feedback data as the predictor variable. The variance normalised to the sum of the variances for (e) the raw data wavelet transformation for the phase (blue) and Rabi (black dashed), and (f) the variance transformed data for the phase predictor (blue) and Rabi predictor (black dashed). (g) The Pearson correlation coefficient squared ($r^2$-value) calculated for each $1/\lambda$ of the wavelet-transformed variables. (g) The peak variance calculated through the variance transformation for all pairs of data. (h-l) Are the same as (c-g) but using the Morlet basis.
}
    \label{fig:variancetransform}
\end{figure}

To demonstrate the variance transformation, we analyze the Q1 phase and Q1 Rabi feedback data, as depicted in Figs.~\ref{fig:variancetransform}(a) and (b), respectively. By applying the variance transformation, described by Eq.~\ref{eq:variance}, to each dataset and examining the outcomes in the Haar basis as shown in Figs.~\ref{fig:variancetransform}(c) and (d), the predictor variable is modified to exhibit spectral features similar to those of the response variable. The variance spectra for each $1/\lambda$ are outlined in Fig.~\ref{fig:variancetransform}(e), with the variance spectra of the variance-transformed datasets showcased in Fig.~\ref{fig:variancetransform}(f). The spectra of the raw data indicate that the majority of noise in both variables is attributable to low $1/\lambda$ (low frequency), with the Rabi variance (black dashed line) also revealing a minor component at approximately $10^{-2}$ s$^{-1}$. After the variance transformation, the variance spectra of the transformed data in Fig.~\ref{fig:variancetransform}(f) suggest the dominant signal in both datasets remains the low $1/\lambda$ noise.

Fig.~\ref{fig:variancetransform}(g) displays the widths at which the datasets undergo the highest variance in the variance transformation using the Haar wavelet. The square sizes reflect the $r^2$-value, while the colour denotes the $1/\lambda$ value. Predominantly, the squares reveal peak variance at lower $1/\lambda$ values, suggesting that the primary high-amplitude noise source correlated within the device exhibits low frequency. Nevertheless, it's essential to recognise that high variance does not equate to high correlation, as the square sizes imply. This insight is critical for pinpointing the most significant noise sources within the system.

Applying the same procedure using the Morlet basis,Figs.~\ref{fig:variancetransform}(h-l), reveals differences between the bases, highlighting the necessity of selecting an appropriate basis for conducting noise analysis in qubit data, and thus emphasising the method's sensitivity to the choice of wavelet basis.

The especially interesting aspect of the variance transformation lies in its capacity to identify the frequencies that hold the high amplitude correlations. This technique aids in determining which variables are most suited for predicting others within a collection of datasets, as demonstrated in Ref.~\cite{Jiang2020RefiningModeling}. Insights of this nature are useful for forecasting device parameters and can aid the refinement of feedback protocols in the context of the data set analyzed in this work.

\section*{Conclusion}
In this work, we adopted Haar and Morlet wavelets, extending their use to qubit noise analysis. Such analysis could refine qubit operations through knowledge of noise sources and forecast outcomes more accurately. Identifying noise origins is vital for improving noise models in solid-state devices, potentially optimizing control pulses and tailoring error correction codes. Scaling quantum computing requires scalable noise analysis and control strategies. Our approach demonstrates scalable noise analysis and feedback optimization based on variable correlations, enhancing quantum computing efficiency. Analysing spatial and temporal noise in a SiMOS two-qubit system can advance design of better error correction codes and error mitigation strategies. Wavelet analysis's ability to detect signal jumps and analyze correlations that occur at specific times offers insights into noise origins in CMOS-based qubit hardware, promising to support scalable qubit system development.

\section*{Methods}

\subsection*{Experimental set up}
To control the qubit device's potentials direct current (DC) and alternating current (AC) signals are combined at room temperature before being routed to the qubit device inside a $^3$He/$^4$He dilution refrigerator. DC signals are generated using SRS SIM928 isolated voltage sources. AC signals up to $\sim 100$ MHz are synthesised using an Quantum Machines OPX system. For spin manipulation we use wideband single-sideband IQ modulation of a Keysight E8267D PSG Microwave Signal Generator with a carrier frequency set around 20 GHz, determined by the global magnetic field. The IQ modulation signals are generated using the same OPX system. With the OPX's FPGA capabilities we are able to change individual qubit control frequencies, phases and control amplitudes in real time. Single electron transistor readout signals are acquired using a Basel Precision instruments SP983c IV converter, SRS SIM910 JFET preamplifier and SRS SIM965 low-pass filter, processed and acquired using the OPX digitisation input. An in-depth discussion on the implementation of the experiment can be found in Ref.~\cite{DumoulinStuyck2023FeedbackExperiment}.  

\subsection*{Discretizing the continuous wavelet transform}
The continuous wavelet transform of a function $x(t) $, where $t $ has units of time (e.g., seconds), is defined as:
\begin{equation}
    W(\lambda, \tau) = \int_{-\infty}^{\infty} x(t) \psi_{\lambda,\tau}^*(t) \, dt,
\end{equation}
where:
\begin{itemize}
    \item $\lambda $ (width) has units of time, inversely related to frequency.
    \item $\tau $ (translation) has units of time, indicating the position of the wavelet.
    \item $\psi_{\lambda,\tau}(t) = \lambda^{-1/2} \psi\left(\frac{t-\tau}{\lambda}\right) $ is the scaled and translated wavelet.
    \item $\psi_{\lambda,\tau}^*(t) $ denotes the complex conjugate of $\psi_{\lambda,\tau}(t) $.
\end{itemize}
The width $\lambda $ adjusts the wavelet to different frequencies, while $\tau $ shifts the wavelet along the time axis, allowing for the analysis of the signal at different times.

To adapt the continuous wavelet transform for digital signal processing, the parameters $\lambda $ and $\tau $ need to be translated into dimensionless indices suitable for discrete data processing. This transition involves:
\begin{enumerate}
    \item \textbf{Sampling}: The continuous signal $x(t) $ is sampled at discrete time intervals $\Delta t $, resulting in a series of data points $x_n = x(n\Delta t) $, where $n $ is an integer.
    \item \textbf{Discretization of Parameters}: The continuous parameters $\lambda $ and $\tau $ are replaced by dimensionless indices $m $ and $k $ respectively. The scale index $m $ corresponds to different widths $\lambda_m = m\Delta t $, and the translation index $k $ represents shifts in terms of data points $\tau_k = k\Delta t $.
\end{enumerate}

The discrete version of the continuous wavelet transform for a digital time series $\vec{x} $ is then defined as:
\begin{equation}
    W(m,k) = \sum_{n=0}^{N-1} x_n \psi_n^*(m,k),
\end{equation}
where $\psi_n^*(m,k) $ is the discretized wavelet function, adapted for the dimensionless indices $m $ and $k $.

The results of the discrete wavelet transform can be organised into a matrix, which represents the wavelet coefficients for each scale and position:
\begin{equation}
\bm{W}_x = 
\begin{pmatrix}
W(0,2) & W(0,4) & \cdots & W(0,K) \\
W(1,2) & W(1,4) & \cdots & W(1,K) \\
\vdots & \ddots & \ddots & \vdots \\
W(N-1,2) & W(N-1,4) & \cdots & W(N-1,K)
\end{pmatrix},
\end{equation}
where each row corresponds to a different time shift and each column to a different scale. This matrix has dimensions $N \times \left(\frac{N-1}{2}\right) $, assuming $K = N-1 $ and that we only consider even $k $.

\subsection*{Comparing different bases}
When the Haar wavelet transformation is applied, it quantifies the difference between two sections of a dataset at a specified width $\lambda$. The Haar wavelet is known for having one vanishing moment, which restricts its capability to encode signals to those that are constant or can be represented by zero-degree polynomials. This attribute makes it particularly effective for encoding constant signals or detecting discrete edges, but not as suitable for capturing signals with more complex variations.

Wavelets with a greater number of vanishing moments have the ability to capture linear or quadratic trends within the data. This leads to a more compact representation with a reduced number of wavelet coefficients $W(\tau,\lambda)$. Such a mechanism is fundamental to the application of wavelets with multiple vanishing moments in fields like image compression algorithms, as detailed in Ref.~\cite{Unser2003MathematicalFilters}.

Conversely, the Morlet wavelet allows for the adjustment of vanishing moments through the parameter $\epsilon$, enhancing its flexibility in feature detection. The condition $\sum_{n=0}^{N-1}\psi_n=0$ requires the Gaussian width parameter $\epsilon$ to exceed 5. This is to prevent the wavelet from reducing to a simple Gaussian function with no oscillations. In this work, we opt for $\epsilon=5$, which offers a balance between capturing oscillatory features and maintaining a practical Gaussian width.

\bibliography{references}

\section*{Acknowledgements}
We acknowledge support from the Australian Research Council (FL190100167 and CE170100012), the US Army Research Office (W911NF-23-10092), and the NSW Node of the Australian National Fabrication Facility. The views and conclusions contained in this document are those of the authors and should not be interpreted as representing the official policies, either expressed or implied, of the Army Research Office or the US Government. The US Government is authorised to reproduce and distribute reprints for Government purposes notwithstanding any copyright notation herein. A.E.S, S.S., and J.Y.H. acknowledge support from the Sydney Quantum Academy. 

\section*{Author contributions statement}
W.H.L. and F.E.H. fabricated the devices, with A.S.D.’s supervision, on isotopically enriched $^{28}$Si wafers supplied by K.M.I.(800ppm). N.D.S., W.G., T.T., J.Y.H., and W.H.L. did the experiments, coding and initial analysis, with A.L., A.S., C.H.Y., and A.S.D.’s supervision. A.E.S. created the correlation analysis methodology with input from S.S., and with A.S.'s supervision. N.D.S. and A.E.S. did the feedback analysis. N.D.S. and A.E.S. wrote the manuscript, with the input from all authors.

\section*{Competing interests}
A.S.D. is CEO and a director of Diraq Pty Ltd. N.D.S., T.T., W.G., F.E.H., A.L., W.H.L., C.H.Y, A.S.D., and A.S. declare equity interest in Diraq Pty Ltd. A.E.S., S.S., J.Y.H and K.M.I. have no competing interests.

\section*{Data availability}
The data that support the findings of this study are available from the corresponding authors upon reasonable request.

\end{document}